# Multibit memory operation of metal-oxide bi-layer memristors

Spyros Stathopoulos[1], Ali Khiat[1], Maria Trapatseli[1], Simone Cortese[1], Alexandrou Serb[1], Ilia Valov[2] and Themis Prodromakis[1*]

[1]Department of Electronics and Computer Science, Faculty of Physical Science and Engineering, University of Southampton, University Road, SO17 1BJ, Southampton, United Kingdom.

[2]Forschungszentrum Jülich, Wilheml-Johnen-Straße, 52428, Jülich, Germany

[*]Corresponding Author: Themis Prodromakis (Email: t.prodromakis@soton.ac.uk)

**In this work, we evaluate a multitude of metal-oxide bi-layers and demonstrate the benefits from increased memory stability via multibit memory operation. We introduce a programming methodology that allows for operating metal-oxide memristive devices as multibit memory elements with highly packed yet clearly discernible memory states. We finally demonstrate a 5.5-bit memory cell (47 resistive states) with excellent retention and power consumption performance. This paves the way for neuromorphic and non-volatile memory applications.**

Emerging memory devices, also known as memristors[1], are nowadays attracting considerable attention due to the breadth of potential applications ranging from non-volatile memory[2] to neuromorphic systems[3,4] and reconfigurable circuits[5]. Their competitive advantage over established complementary metal-oxide-semiconductors (CMOS)-based memory stems from their capability to support a multitude of states, long retention characteristics, fast switching and ultra-low power consumption[6]. Many technologies have been put forward as potential winners of the non-volatile memory race[7], namely phase-change (PCRAM)[8], magnetic (MRAM)[9] and resistive random access memory (ReRAM)[10]. Although MRAM and PCRAM are considered more reliable, they are constraint by power and/or scalability issues[11,12]. In contrast, ReRAM has shown capacity of operating in the femtojoule regime[13], with functional devices reported at feature sizes that outperform CMOS[14,15]. Although the realisation of bistable memory devices (1-bit) is apparent from the very nature of the memristor to variate between two resistive states[16] the implementation of a device that can reliable be programmed at a multitude of distinct resistive states still poses a significant challenge.

Resistive switching has been observed in many metal-oxide systems[17], with $Ta_2O_5$[18,19], $HfO_2$[20] and $TiO_2$[21,22] being among the most popular. In all cases, the origin of switching has been attributed to either the drift of oxygen vacancies[22] and/or interstitials[23] or the formation of conductive filaments[24] within an active metal-oxide core under the influence of an applied field. Studies have reported that the introduction of a thin interfacial barrier layer between the active layer and one of the electrodes can influence the devices' stability and/or reduce the power consumption of $TiO_2$ ReRAM[25-28].

We prepared samples with seven different active layer configurations and platinum top and bottom electrodes: 1) $TiO_2$-only; 2) $Al_xO_y/TiO_2$; 3) $Ta_xO_y/TiO_2$; 4) $SiO_2/TiO_2$; 5) $ZnO/TiO_2$; 6) $HfO_x/TiO_2$ and 7) $WO_x/TiO_2$. The thickness is 4 nm and 40 nm for the barrier layer and $TiO_2$ respectively. The dimensions of the active area of the devices were 20×20 μm$^2$. Using 1 μs pulses of negative polarity ranging from -3 to -12 V with 100 mV step devices were electroformed to a usable resistance range, typically 20–150 kΩ depending on the stack (see supplementary fig. S1 for typical electroforming behaviour). Figures 1d and 1e show a comparison between the $TiO_2$-only and $Al_xO_y/TiO_2$ stacks in respect to the pulsing stability of the fabricated devices. 100 ns pulse ramps of alternating polarity from 1 to 2 V with 200 mV step have been applied to both devices. Considerable drift is apparent in the $TiO_2$-only devices (fig. 1d) which practically eliminates the memory window after 20 switching cycles. Although the stability of $TiO_2$-only device can be further optimised at the expense of energy (see supplementary fig. S2) the $Al_xO_y/TiO_2$ is clearly more stable (fig. 1e) exhibiting sharply defined high and low resistance regions and maintaining a constant OFF/ON ratio throughout the experiment.

The effect of reliable pulsed switching becomes more apparent in their multibit capabilities. Devices have been evaluated for their multistate performance by biasing them with 100 ns pulses ranging from 1 to 2 V at 50 mV step. A new state is assumed if it is stable and the lower bound of the standard deviation of a series of 50 read pulses (at 0.5 V) is at least 2σ higher than the upper bound of the previous state (see Methods and supplementary figures S3–S5 for more detail). Using this evaluation routine we observed a significant increase in the number of attainable resistive states for the bilayer devices in contrast to the monolayer cells. While in the case of $TiO_2$-only devices only a maximum of 10 states was observed the introduction of a barrier layer both increased the number of resistive states significantly and improved the dynamic response of the devices. In fig. 2 a comparison between the different layer combinations in respect to the number of attainable states and the final increase over the initial resistive state can be observed. All combinations resulted in an increased number of resistive states and dynamic range.

Although all the barrier layer combinations gave us improved results in multibit capability, the $Al_xO_y/TiO_2$ combination proved to yield the highest "state expanse" ($\max\{R/R_o\} \times$ (# of states), where $R_o$ is the baseline resistance). Figure 3 shows the result of multibit characterisation for those devices. Because of the improved characteristics of the $Al_xO_y/TiO_2$ stack it is possible to arbitrarily program the device to a specified resistive state. As can be seen in Fig. 3a using singular 100 ns SET pulses at 2 V allows us to sequentially select the resistive state of the device. Each pulse raises the resistance of the device to a well-defined value. Selection of a different state can be done by "flushing" the device back to its baseline resistance of ~21.5 kΩ with a train of 100 ns RESET pulses at −2 V and then applying a different number of SET pulses. Despite the continuous SET/RESET cycles both the baseline resistance as well as the individual resistive states remain stable and reproducible. This is in effect a proof-of-concept programming protocol for the implementation of a non-volatile RRAM-based random access memory (NVRAM) cell.

A maximum of 47 stable resistive states are reported which allows to establish *a new state of the art* figure for multibit non-volatile information storage at 5.5 bits per cell with an average step of ~1.2 kΩ step per state. The cumulative probability distribution function graph (fig. 3c) clearly illustrates the overall discernibility of all the resistive states. Retention characteristics of select states (fig. 3d) over a period of 8 hours are also excellent with only the higher resistive states that are close to the volatility threshold significantly observable fluctuation. Resistive states are stable and even in the 30–40 kΩ range where the states are closely packed they remain clearly distinguishable.

The excellent performance and stability of states of bilayer structures can be attributed to the specific ratio of the ionic transference numbers of the second oxide layer. Looking at fig. 2 there is a clear trend for the number of available states, whereas no particular trend on particular dependence can be observed. The highest number stable multilevel states is achieved with $Al_2O_3$, followed by $Ta_2O_5$, $WO_3$, $HfO_2$, ZnO and $SiO_2$. It has been recently shown that many oxide thin films used for RRAMs have mobile host cations [24] and that obviously the oxidation state and stoichiometry of the matrix is also playing a significant role [25]. Mobility of cations and anions during high field oxide formation on metals using liquid electrolytes is known from the classical electrochemistry. In high voltages and low film thicknesses conditions the transport is field accelerated and the particular ionic transference numbers depend on the field strength and the current density. The metaloxide with highest cation cation transference number is $Al_2O_3$, followed by $Ta_2O_5$, $WO_3$ and $HfO_2$ [26–28]. The order shown in fig. 2 strictly correlates with the higher mobility of cations or lower mobility of oxygen ions, respectively. Similar effect of the oxygen mobility on the device stability has been reported for STO using barrier layers of $Al_2O_3$ (low $O_2^-$ mobility) and yttria-stabilized $ZrO_2$ (high $O_2^-$ mobility) [29]. Thus we can conclude that the main factor influencing our device performance is the transport properties of the film added to the $TiO_2$ layer.

It is important to mention that our characterisation routine foregoes the use of compliance current limiting to switch the device to a higher (or lower) resistance. Current compliance limiting is a common practice that is used to control the size of the conductive filament and consequently the resistance of the device [43,44]. Instead we have opted for a more direct approach by sequentially pulsing the device until its state stabilises. As the energy budget is incrementally increased until the resistance exceeds a predefined tolerance we ensure that the minimum amount of required switching energy is expended. Figure 3e depicts the calculated energy requirements to attain any of the 47 states of the $Al_xO_y/TiO_2$ device. In order to calculate the energy evolution the formula $\sum\{V^2 / R_{min,max} \Delta t\}$ has been applied for each resistive state, where V is the pulsed voltage level and Δt the pulse width. Since biasing is always between 1 and 2 V $R_{\mathrm{min,max}}$ represent the resistance in these two voltages as calculated from the I–V characteristic (see also figure S8). For all the states of the $Al_xO_y/TiO_2$ device the switching energy remains in the pJ–nJ range.

In this paper we presented the realisation of a state of the art 5.5-bit ReRAM device cell for non-volatile memory applications. Using a bilayer device stack and a novel current compliance-free characterisation protocol we managed to achieve 47 stable resistive states as well as an overall improvement on the reliability compared to the $TiO_2$-only devices. This achievement can establish bilayer-based memristors as a viable technology path for the implementation of next generation non-volatile memory devices and neuromorphic applications.

## Methods

**Device fabrication: (150 words ~ 9 lines):** All devices have been fabricated on 6-inch oxidised silicon wafers (200 nm of thermal $SiO_2$). Initially the bottom electrodes were fabricated using photolithography and electron beam evaporation of titanium (5 nm) and platinum (10 nm) followed by lift-off process in N-Methyl-2-pyrrolidone (NMP). Then, 45 nm of $TiO_2$ were deposited using magnetron sputtering. The $Al_2O_3$, $Ta_2O_5$, and $SiO_2$ layers (4 nm) were also deposited using magnetron sputtering after negative tone photolithography. The active layer is formed after lift-off in NMP. The 4 nm layers of ZnO, $HfO_2$ and $WO_3$ were synthesised using atomic layer deposition (ALD). After that a positive tone photolithography and ion beam milling processes were used to pattern and etch the active layers. The top electrode was fabricated using photolithography, electron beam evaporation of platinum (10 nm) and lift-off in NMP.

**Electrical Characterisation: (150 words max):** Characterisation of the memristors has been done with our in-house memristor characterisation platform [30]. Devices are initially electroformed to a usable resistance range (25 to 200 kΩ, depending on the stack) using consecutive 1 μs pulses of negative polarity ranging from -8 to -12 V in amplitude. A series resistor of 1 kΩ was used as a current-limiting mechanism for all devices. Resistance initially drops to the $10^6$ Ω range and then to a more stable $10^4$–$10^5$ Ω range. Multi-bit capability of the devices has been evaluated with a custom algorithm (see following section). In order to extract the retention curve a sequence of 100 ns 2 V pulses is used to program the device to a specified resistance and then a read pulse is applied every 5 minutes for 8 hours.

**Resistive state evaluation algorithm (100 words):** State assessment occurs over three phases. During the first phase a series of programming pulses of a predefined duration (100 ns), increasing amplitudes and alternating polarities is applied to the device under test and the resistive state of the device is evaluated between every pair of programming trains. This is to determine the polarity that induces a switch in the resistance of the device. After the switching polarity has been determined the second phase, using fixed amplitude, 100 ns pulses of the opposite polarity in respect to the one determined in the first phase, drives the resistance to a stable low value. Stability is assumed when the fitted slope is lower that a predefined threshold. The third phase applies an increasing number of 100 ns programming pulsing using the polarity determined from the first phase followed by two read trains separated by a 100 ms retention interval. If the lower bound of the standard deviation of the resistance measured between these trains is at least 2σ higher than the upper bound of the previous state a new resistive state is established. The algorithm terminates if the voltage limit is reached or if the trend of the resistive states become non-monotonic. The granularity on the standard deviation directly impacts the number of assessed states (see supplementary figure S3). 2σ was used throughout the electrical characterisation as it provides a large enough confidence interval (at least 95%) while allowing the exploitation of a high amount of resistive states. A flowchart detailing the steps of the algorithm described here can be found in supplementary Fig. S4.

**Data Availability:** The data that support the findings of this study are available from the corresponding author upon request, as detailed in http://www.nature.com/authors/policies/data/data-availablity-statements-data-citations.pdf.

**Supplementary Information** is available in the online version of the paper.

**Acknowledgements** We acknowledge the financial support of FP7 RAMP and EPSRC EP/K017829/1.

**Author contributions** S.S and A.K contributed equally to this work. T.P, A.K and S.S conceived the experiments. A.K and M.T optimized the fabrication process flowchart and fabricated the devices. S.C performed the preliminary measurements. S.S and A.S developed the algorithm. S.S, A.K and T.P performed and optimized the electrical characterisations. I.V wrote the mechanism section. S.S, A.K and T.P wrote the manuscript. All authors contributed in improving writing the manuscript.

**Author Information** Reprints and permissions information is available at… The author declare no competing financial interests. Correspondence and requests for materials should be addressed to T.P. (t.prodromakis@soton.ac.uk)

## Figure captions

**Fig. 1: Comparison between $TiO_2$-only devices and $Al_xO_y/TiO_2$ bilayer devices** (a) SEM micrograph of a memristor device; (b) Schematic representation of a single layer $TiO_2$-based device with platinum top and bottom electrodes; (c) Schematic representation of a bilayer $Al_xO_y/TiO_2$-based device with platinum top and bottom electrodes; (d) Typical bipolar switching of a device based on the stack pictured in (b) using 100 ns pulses of alternating polarity voltage ramps ranging from 1 to 2 V, with voltage steps of 200 mV; (e) Typical bipolar switching of a device based on the stack pictured in (c) using 100 ns pulses of alternating polarity voltage ramps ranging from 1 to 2 V with voltage steps of 200 mV. The coloured horizontal lines in fig. (d) and (e) denote the average low (LRS) and high resistive state (HRS).

**Fig. 2: Multibit evaluation of devices based on different barrier layer combinations.** Number of attainable resistive states (left axis) and ratio of the final state resistance over the baseline resistance (right axis). Confidence interval for the state assessment is 2σ following the routine described in supplementary Fig. S4 and S5. A chart containing each individual state assessed for every bilayer combination can be found in supplementary Figs. S6 and S7.

**Fig. 3: Multibit operation of a device using the $Al_xO_y/TiO_2$ RRAM stack.** (a) Arbitrary programming of specific resistive states. Starting from a baseline resistance of ~21.5 kΩ and using sequential 100 ns SET pulses at 2 V, as shown in (b), the device can be programmed accurately to a specific state. In order to switch into a different state the device is “flushed” using a train of 100×100 ns RESET pulses at −2 V resetting it to its baseline resistive state; (c) Cumulative probability

distribution function plot for each of the established resistive states. All states are closely packed and individually discernible. (d) 8-hour retention measurements for select resistive states. After the resistive state assessment the device is driven back to its baseline resistance using 100 ns −2 V pulses. (e) Switching energy required to attain the resistive states shown in figure (c). There is an exponentially increasing energy requirement in order to achieve higher resistive states.

## Fig. 1

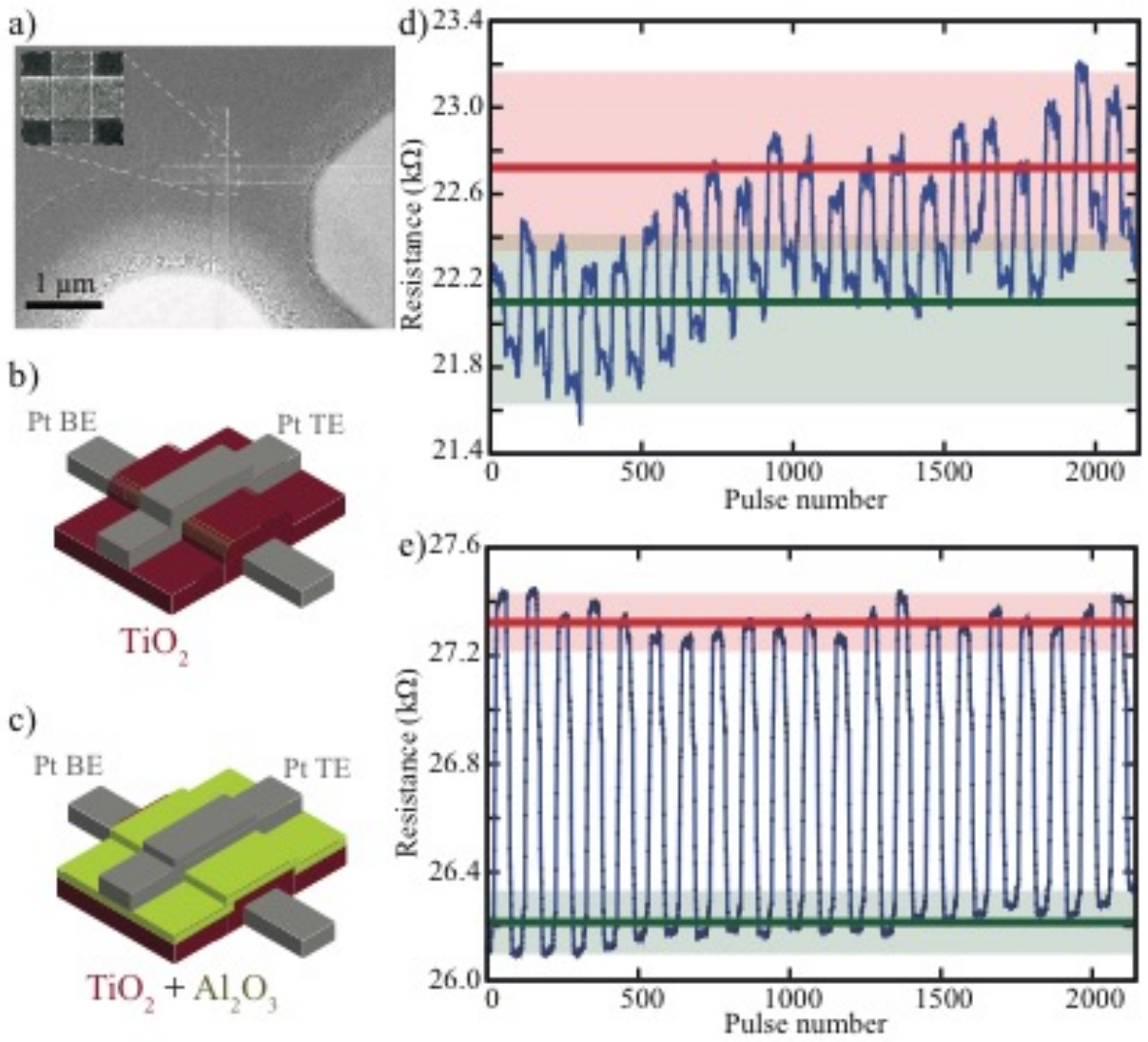

## Fig. 2

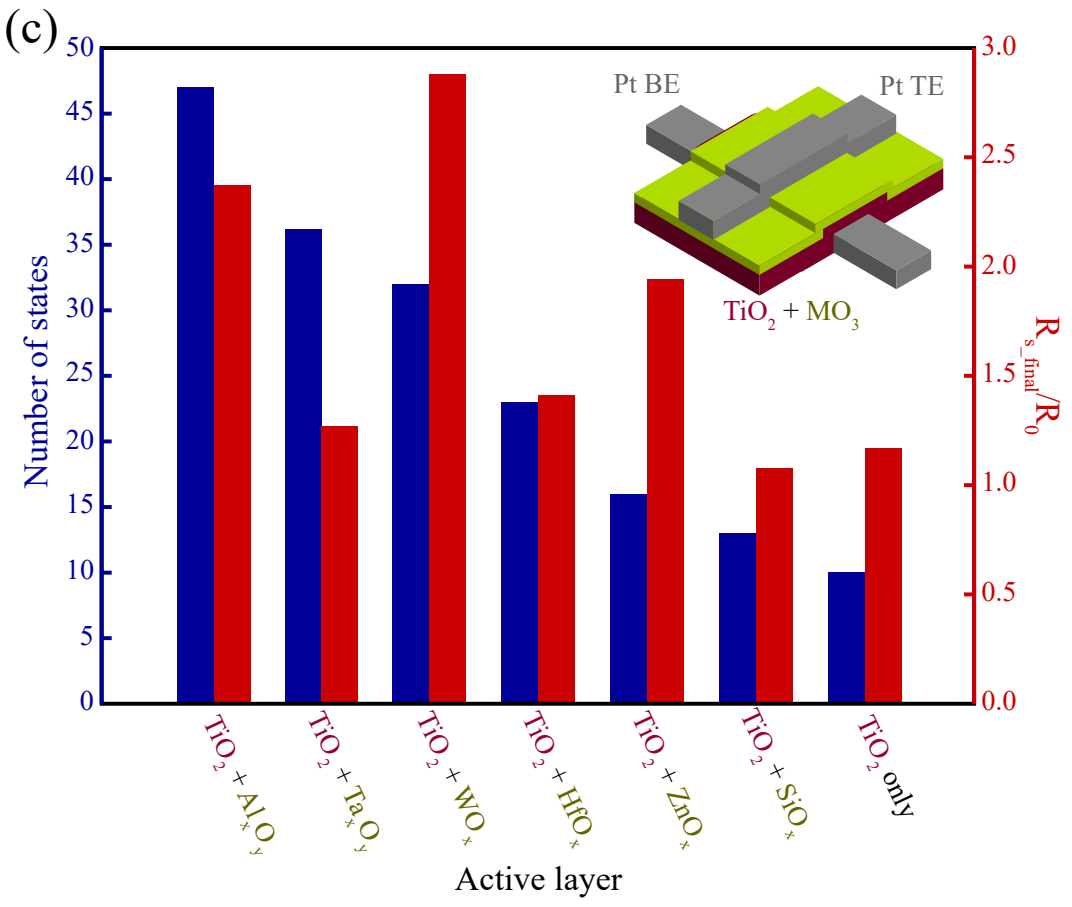

## Fig. 3

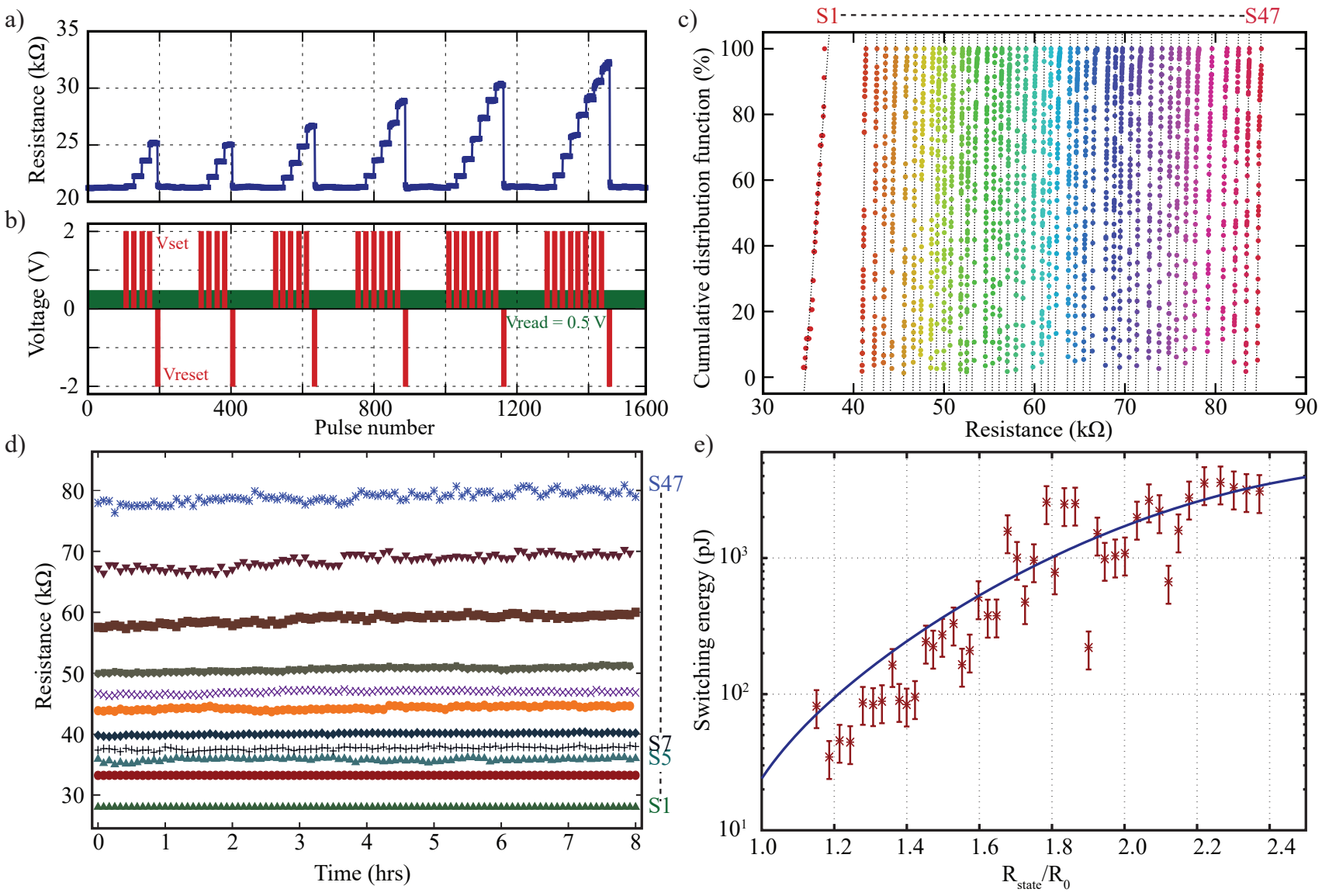

## Supplementary Information

**Supplementary Fig. 1: Electroforming behaviour of an $Al_xO_y/TiO_2$ device.** Electroforming was performed using 1 μs pulses ranging from -3 to -12 V for all devices. A typical response to this electroforming protocol, exhibited here in an $Al_xO_y/TiO_2$ device, is an initial drop in the resistance (here at ~75 kΩ) at around 10 V followed by a further drop into the usable initial resistance range.

**Supplementary Fig. 2: Optimised pulsing protocol for a $TiO_2$-only device.** Stability of the $TiO_2$-only can be improved when using ramps of 1 μs voltage pulses ranging from 1 to 3 V with 100 mV step and alternating polarities. Although the required energy is increased (longer pulses, higher voltage) the stability of the $TiO_2$-based device improved to the point that the worst-case switching windows between low and high resistive states are non-overlapping.

**Supplementary Fig. 3: Assessment between two adjacent resistive states for a device using the $Al_xO_y/TiO_2$ stack.** An increasing number (up to 10) of 100 ns programming pulses (b) is applied with 50 mV step. In-between the programming pulses there are 50×0.5 V read pulses. During the last 50 read pulses the lower bound of resistance of the device should be at least 2σ greater that the upper bound of the resistance of the previous state (51.86 ± 0.17 kΩ in the shown example) and therefore a new resistive state is established at 52.98 ± 0.14 kΩ.

**Supplementary Fig. 4: Block diagram of the state evaluation algorithm.** State evaluation happens over three distinct phases. (a) *Phase I* determines the switching polarity of the device by applying pulses of alternating polarity. If the applied pulse causes a resistance response outside a predefined tolerance band the switching polarity S is determined to be either positive (if the final resistance is above the tolerance band) or negative (if the final resistance is below the tolerance band). (b) *Phase II* drives the resistance of the device to a stable low or high level. Depending on the outcome of Phase I a series of pulses of opposite polarity is applied until the the slope of the fitted resistance response is less that a predefined threshold. A minimum of 50 points is accumulated for this evaluation. (c) *Phase III*: initially a base resistance and its standard deviation is calculated. This calculation is composed of two sets of 25 read pulses separated by a (configurable) retention time of 100 ms. Afterwards a train of pulses of constant voltage and width is applied using the polarity determined from Phase I. The resistance of the device is evaluated again using the same method as the one used for the base resistive state. A new state is established if the lower/upper bound of the standard deviation of the new resistive state is at least 2 or more standard deviations above the upper/lower bound of the standard deviation of the previous resistive state. Otherwise the voltage is increased and the process repeats. The algorithm terminates if a maximum voltage is reached or the resistive state sequence becomes non-monotonic

**Supplementary Fig. 5: Effect of the confidence bounds in the number of attainable states of the $Al_xO_y/TiO_2$ device.** The maximum number of possible attainable states depends on the confidence bounds used. By using 3σ (99.7%) instead of 2σ (95%) the number of registered resistive states is roughly halved (23 from 47). For most practical scenarios, however, a 95% confidence interval is sufficient to discern two adjacent states.

**Supplementary Fig. 6: "Short term" retention for all bilayer combinations.** The results of the final state measurements (50×100 ns pulses with 20 ms interval). With the exception of $SiO_2$ all bilayer combinations improve both the number of attainable states and the overall stability of each established state in comparison to the $TiO_2$-only device.

**Supplementary Fig. 7: Multistate evaluation for different bi-layer combinations.** The chart depicts all established resistive states with 2σ confidence for each bilayer combination studied in this paper. All combinations are improving the $TiO_2$ stack regarding the number of states but only $Al_xO_y/TiO_2$, $WO_3/TiO_2$ and $HfO_2/TiO_2$ stacks also provide an increase the dynamic range of the device. The 46 resistive states, the overall linearity as well as the improved dynamic range constitute the $Al_xO_y/TiO_2$ the most promising combination for granular, predictable, multi-bit storage.

**Supplementary Fig. 8: I–V characteristic of an $Al_2O_3/TiO_2$ device in the ~25 kΩ range.** In order to calculate the required switching energy (see main text fig. 4b) the formula $\Sigma\{V^2/R_{min,\,max}\cdot\Delta t\}$ was used. $R_{min}$ and $R_{max}$ are extracted by multiplying the resistance at READ voltage (0.5 V) with the ratio of the slopes at 0.5 V versus 1 V (for Rmin) or 2 V (for Rmax). This is a very conservative "worst case" approach to estimate the energy usage for the device. Starting from the low resistive state, which is the most energy consuming state, we take a current–voltage characeteristic that covers the relevant switching range. Since we are starting from the base resistive state of the device the I–V shows no signs of further setting even when the voltage is up to 2 V therefore power dissipation for a device subjected to 2 V and setting must be necessarily lower that our assessment based on this I–V.